\documentclass[useAMS]{mn2e}
\usepackage[dvips]{graphicx}
\usepackage{color}
\usepackage{amssymb}

\title[Carbon abundance in A-, F- and G-type supergiants]
{\uppercase{
	Carbon abundance and the N/C ratio in atmospheres of A-, F- and G-type supergiants and bright giants}}
	
	\author[L.S. Lyubimkov et al.] 
{Leonid~S.~Lyubimkov,$^1$\thanks{E-mail: lyub@crao.crimea.ua (LSL); dll@astro.as.utexas.edu (DLL);
} David~L.~Lambert,$^2$$^\star$ Sergey A. Korotin,$^3$ \and Tamara M. Rachkovskaya,$^1$  Dmitry~B.~Poklad,$^1$ \\
$^1$Crimean Astrophysical Observatory, Nauchny, Crimea, 298409, Russia\\
$^2$The W.J.McDonald Observatory of the University of Texas at Austin, TX 78712-0259, USA\\
$^{3}$ Astronomical Observatory of the Odessa National University, Shevchenko Park, Odessa 65014, Ukraine \\} 

\begin{document}

\date{Accepted. Received ; in original form}

\pagerange{\pageref{firstpage}--\pageref{lastpage}} \pubyear{}

\maketitle

\label{firstpage}

\begin{abstract}
Based on our prior accurate determination of fundamental parameters for 36 Galactic A-, F- and G-type supergiants and bright giants (luminosity classes I and II), we undertook a non-LTE analysis of the carbon abundance in their atmospheres. It is shown that the non-LTE corrections to the C abundances derived from C~I lines are negative and increase with the effective temperature $T_{\rm eff}$; the corrections are especially significant for the infrared C~I lines with wavelengths 9060-9660~\AA. The carbon underabundance as a general property of the stars in question is confirmed; a majority of the stars studied has the carbon deficiency [\textit{C/Fe}] between -0.1 and -0.5~dex, with a minimum at -0.7~dex. When comparing the derived C deficiency with the N excess found by us for the same stars earlier, we obtain a pronounced N vs. C anti-correlation, which could be expected from predictions of the theory. We found that the ratio [\textit{N/C}] spans mostly the range from 0.3 to 1.7~dex. Both these enhanced [\textit{N/C}] values and the C and N anomalies themselves are an obvious evidence of the presence on a star's surface of mixed material from stellar interiors; so, a majority of programme stars passed through the deep mixing during the main sequence (MS) and/or the first dredge-up (FD) phase. Comparison with theoretical predictions including rotationally-induced mixing shows that the stars are either post-MS objects with the initial rotational velocities $V_0$~= 200-300 km/s or post-FD objects with $V_0$~= 0-300 km/s. The observed N vs. C anti-correlation reflects a dependence of the C and N anomalies on the $V_0$ value: on average the higher $V_0$ the greater the anomalies. It is shown that an absence of detectable lithium in the atmospheres of the stars, which is accompanied with the observed N excess and C deficiency, is quite explainable.
\end{abstract}


\section{Introduction}

      Carbon and nitrogen are two key light chemical elements, whose atmospheric abundances can significantly change during the stellar evolution. Therefore, carbon and nitrogen provide an opportunity for testing modern evolutionary theories. We consider here stars with masses \textit{M} between 4 and 25 M$_\odot$, which are observed at first as early and middle B-type main sequence stars (luminosity classes V, IV and III). As known, the CNO-cycle is a main source of energy in the stars with such \textit{M} during this evolutionary phase; this cycle results in significant changes in the C and N abundances inside the stars. Next these stars are observed as A, F and G-type supergiants and bright giants (luminosity classes I and II). The latter objects are the focus of the present paper.

      Luck \& Lambert (1985) showed that non-variable F, G, and early K supergiants reveal systematic carbon underabundance, as well as systematic nitrogen overabundance, i.e. nitrogen and carbon abundances were anti-correlated (see, e.g., Lyubimkov's 1998 review). The anti-correlation was predicted by the theory of stellar evolution because the supergiants develop a convective envelope which brings to the surface interior material exposed to the H-burning CNO-cycle which processes carbon to nitrogen. Such N and C anomalies were considered initially as solely the result of the deep convective mixing during the red giant/supergiant evolutionary phase (known as the first dredge-up, hereinafter FD). Now it is known that the rotationally-induced mixing during the earlier main sequence (MS) evolutionary phase can play an important role in appearance of these anomalies, too. In the present paper we compare the observed C and N anomalies of A-, F- and G-supergiants with theoretical predictions for two evolutionary phases: (i) directly after the termination of the MS phase (post-MS phase); and (ii) near the end of the first dredge-up (post-FD phase).

      The sample of AFG supergiants and bright giants has been studied by us earlier in a series of papers providing: (a) accurate determination of fundamental parameters for 63 stars, including their effective temperature $T_{\rm eff}$, surface gravity $\log g$, microturbulent velocity $V_t$ and index of metallicity [\textit{Fe/H}] (Lyubimkov et al. 2010, Paper I); (b) the non-LTE analysis of the nitrogen abundance for 30 stars that confirmed the N enrichment in atmospheres of this type stars (Lyubimkov et al. 2011, Paper II); (c) the non-LTE analysis of the lithium abundance for 55 stars which led to some interesting conclusions after comparison with theory (Lyubimkov et al. 2012, Paper III).

      We present here our results of a non-LTE carbon abundance determination for 36 Galactic A-, F- and G-type supergiants and bright giants (luminosity classes I and II). We used in this analysis the parameters $T_{\rm eff}$, $\log g$, $V_t$ and [\textit{Fe/H}] found in Paper I. The derived C abundances are compared with the N abundances from Paper II; the existence of the C vs. N anti-correlation is confirmed. The C abundance, as well as the N/C ratio (a very sensitive indicator of stellar evolution) as a function of the effective temperature $T_{\rm eff}$, surface gravity $\log g$ and mass \textit{M} is considered. A comparison of the observed C and N anomalies with theoretical predictions is presented. The C and N anomalies in comparison with the Li abundances from Paper III are discussed, too.

\section{SPECTRAL OBSERVATIONS AND THE LIST OF C I LINES}

      We base the present work, like Papers I, II and III, on the high-resolution spectral observations of Galactic AFG-type supergiants and bright giants, which were acquired by us at the McDonald Observatory in 2003-2006 using the 2.7-m telescope and the Tull echelle spectrometer (Tull et al. 1995). Additional observations of some stars were performed in 2009 with the same spectrometer. The stars were observed at a resolving power of \textit{R}~= 60000. The Tektronix CCD with 24-$\rm \mu m^{2}$ pixels in a 2048$\times$2048 format was used as a detector. The spectral region from about 4000 to 9700~\AA\ was covered. The typical signal-to-noise ratio of the extracted one-dimensional spectra was between 100 and 400. Two spectra for each programme star have been obtained.

\begin{table}
 \centering
 \caption{List of the used C I lines}
  \begin{tabular}{|c|c|c|}
  \hline
Line,~\AA\ &Exitation  &$\log gf$\\
&potential,&\\
&eV&\\
\hline
5052.17&7.685&-1.302\\
5380.34&7.685&-1.615\\
5800.23&7.946&-2.802\\
5800.60&7.946&-2.336\\
6001.13&8.643&-2.060\\
6010.68&8.640&-1.937\\
6014.84&8.643&-1.583\\
6413.55&8.771&-2.000\\
6587.61&8.537&-1.002\\
6655.52&8.537&-1.998\\
6671.85&8.851&-1.650\\
7087.83&8.647&-1.441\\
7100.12&8.643&-1.469\\
7108.93&8.640&-1.593\\
7111.47&8.640&-1.084\\
7113.18&8.647&-0.772\\
7115.17&8.643&-0.933\\
7115.19&8.640&-1.472\\
7116.99&8.647&-0.906\\
7119.66&8.643&-1.147\\
7473.31&8.771&-2.042\\
7476.18&8.771&-1.573\\
7483.45&8.771&-1.371\\
7848.24&8.848&-1.730\\
7852.86&8.851&-1.681\\
7860.89&8.851&-1.148\\
8335.15&7.685&-0.436\\
8727.13&1.264&-8.140\\
9061.43&7.483&-0.346\\
9062.47&7.480&-0.454\\
9078.28&7.483&-0.580\\
9088.51&7.483&-0.429\\
9094.83&7.488&0.151\\
9111.80&7.488&-0.296\\
9405.73&7.685&0.286\\
9603.03&7.480&-0.895\\
9658.44&7.488&-0.279\\
   \hline
   \end{tabular}
\end{table}

\begin{figure*}
\centering
\includegraphics[width=95mm, angle=-90]{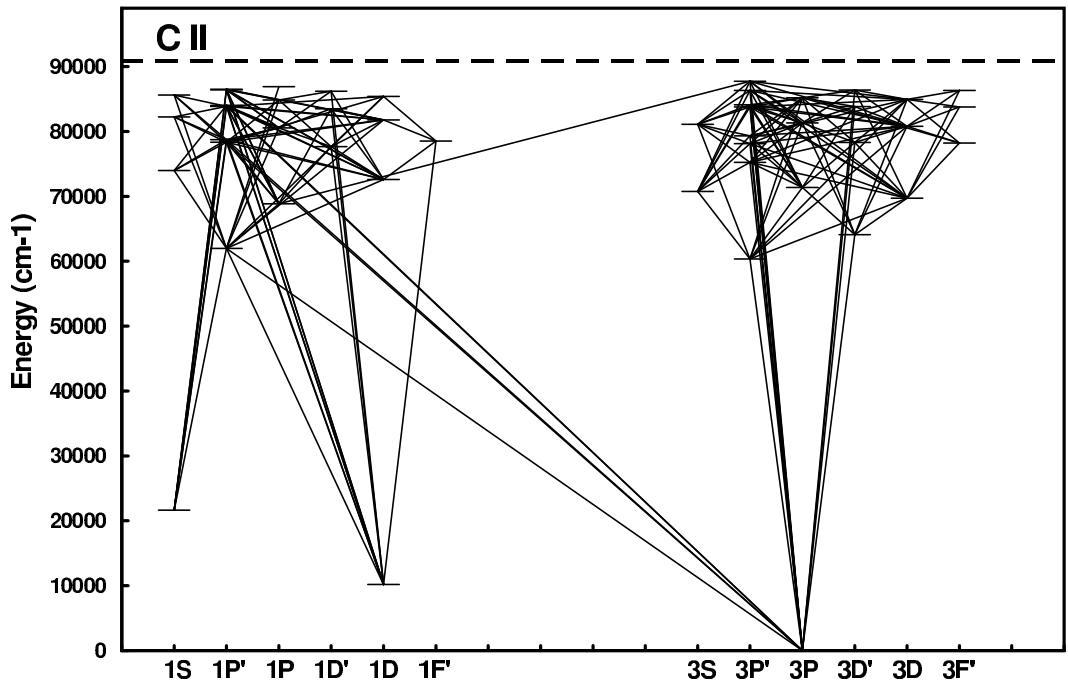}
\caption[]{Model carbon atom.}
\end{figure*}

     Reductions of the observed spectra were performed using the standard IRAF packages. The data reduction includes the following procedures: overscan correction, bias averaging and correction, scattered light subtraction, flat-field combination and normalization, division by flat-field and thorium-argon spectrum wavelength calibration.

      We selected for a detailed analysis 37 C I lines of various multiplets; they are listed in Table 1. We included in the list not only the C I lines in the visual region, but also the lines in the near infrared region up to 9658~\AA. Moreover, the forbidden line at 8727~\AA\ was included, which is interesting due its weak sensitivity to non-LTE effects (it should be noted that this line showed good agreement in the C abundance with other lines). We present in Table 1 the line wavelength in \AA, excitation potential in eV and oscillator strength \textit{gf} according to the VALD database (Kupka et al. 1999; Heiter et al. 2008). The \textit{gf}-value for the forbidden line at 8727~\AA\ is taken from the NIST database (http://physics.nist.gov/PhysRefData/ASD/index.html).

      The problem of line blending is rather important for some C I lines. Therefore, our analysis of these lines is based on computations of synthetic spectra and their comparison with observed spectra. Using the modified code MULTI we calculated the non-LTE populations (and corresponding \textit{b}-factors) for C I levels and then used them in the synthetic spectra calculations with the code STARSP (Tsymbal 1996). These calculations included all spectral lines from the VALD database in a region of interest. The LTE approach was applied for lines other than the C I lines. Abundances of corresponding elements were adopted in accordance with the [\textit{Fe/H}] value for each star. A presence of telluric lines in the observed spectra is taken into consideration, too. 

\section{NON-LTE COMPUTATIONS OF C I LINES}

      Departures from LTE in C I lines for A-, F-, and G-type stars have been analyzed previously; see, e.g., St\"{u}renburg \& Holweger (1990), Takeda (1992), Rentzsch-Holm (1996), Takeda \& Honda (2005), Fabbian et al. (2006). It was found there that non-LTE effects in C I lines are relatively small for dwarfs, like the Sun, but they increase as the effective temperature $T_{\rm eff}$ increases and the surface gravity $\log g$ decreases. Departures from LTE can be significant for A, F and G supergiants, i.e. for the stars of interest here. Therefore, our analysis of C I lines is based completely on the non-LTE computations of these lines. 

      The carbon model atom used in the computations was originally suggested by Andrievsky et al. (2001); now it is updated with new collision rates for C I and C II. The model atom includes 47 levels of C I, as well as 11 levels of C II and a ground level of C III. Moreover, additional atomic levels with LTE populations are included in the equation of the particle number conservation, namely 63 C I levels and 9 C II levels. Excitation energies were taken for C I and C II terms from Moore (1970). Fig.1 shows schematically the adopted model atom (the C II levels are not displayed); only those transitions are presented there that are included in the detailed consideration. 

      Photoionization rates for all included levels were calculated with the help of the photoionization cross-sections from OP TopBASE (Cunto \& Mendoza 1992; Cunto et al. 1993). We selected for detailed consideration 189 radiative bound-bound transitions for C I, and 18 transitions for C II. Radiative rates of other 128 very weak transitions were adopted to be fixed and calculated in the LTE approximation. Oscillator strengths for permitted transitions in question were taken from OP TopBASE. Oscillator strengths for forbidden transitions were taken from Hirata \& Horaguchi's (1994) catalogue. 

      Electron collisional excitation rates for transitions between first 12 levels of C I are based on the effective collision-strengths, which have been computed by Martin O'Mullane and contained in the database ADAS (Summers 2004). Note that O'Mullane's values are in good agreement with results of other authors (Zatsarinny 2005; Suno \& Kato 2006). Effective collision strengths for electron impact excitation for all C II levels were taken from Wilson, Bell and Hudson (2005). For all other allowed bound-bound transitions we used van Regemorter's (1962) formula. Collisional rates for forbidden transitions were calculated with the help of semi-empirical formula (Allen 1976) with a collisional force of 1. The electron collisional ionization was considered using the formula of Seaton (1962) with cross-sections at the ionization threshold from OP TopBASE. Collisions with hydrogen atoms for permitted transitions were introduced according to Steenbock \& Holweger (1984) with a scaling factor 0.4 (see below). 

      For non-LTE computations of C I lines, we used the code MULTI (Carlsson 1986) as modified by Korotin, Andrievsky \& Kostynchuk (1998). Since we apply model atmospheres calculated with the code ATLAS9 (Kurucz 1993), the opacity sources for MULTI are taken from ATLAS9, too. In the modified MULTI code, the mean intensities (they are needed to obtain the radiative photoionization rates) are calculated for a set of frequencies at each atmospheric layer; then they are stored in a separate block, from where they can be interpolated. Castelli's (2005) tables for the opacity distribution function (ODF) are utilized there.  

      When a joint solution of the equations of statistical equilibrium and radiative transfer has been obtained, terms were split into several sublevels whose populations were determined in accordance with their statistical weights. Then the profile of a C I line was calculated. 

       In order to test the adopted carbon model atom, as well as our non-LTE technique of the C I line analysis, we performed a comparison of our computed line profiles with the observed high-resolution spectra of the Sun and the well-studied nearby F5-type dwarf Procyon which has the solar-like chemical composition. We used for the Sun the Solar Flux Atlas of Kurucz et al. (1984); the solar photospheric model was taken from Castelli (http://wwwuser.oats.inaf.it/castelli/sun/ ap00t5777g44377k1asp.dat) with a chromospheric contribution from the VAL-3C model of Vernazza et al. (1981) and the corresponding microturbulence distribution. The observed spectrum of Procyon was taken from the database UVES POP (Bagnulo et al. 2003). 

      These test computations showed that non-LTE effects are different for various C I multiplets. Departures from LTE seem to be small for the C I lines in the visual region; the computed lines are somewhat strengthened, that leads to the non-LTE corrections about -0.05 dex in the derived C abundances. However, already for the lines in the near IR region (7087--7119~\AA) the non-LTE effects become slightly stronger, and for the lines of multiplets No. 2, 3 and 9 (9603--9658, 9061--9111, and 9405~\AA) they lead to increases of computed equivalent widths by 20--25 per cent and, accordingly, to the non-LTE corrections up to -0.25 dex in the C abundances. 

\begin{figure}
\begin{center}
\includegraphics[width=83mm]{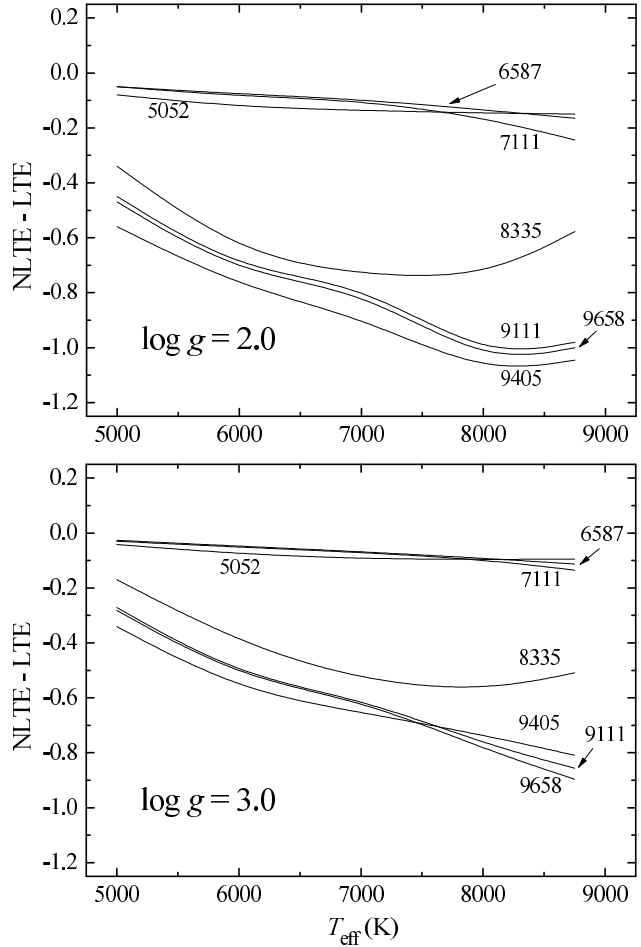}
\end{center}
\caption[]{Non-LTE corrections to the LTE carbon abundances calculated for a number of C I lines from 5052 to 9658~\AA\ as a function of $T_{\rm eff}$. Upper and lower panels correspond to the surfaces gravities $\log g$~= 2.0 and 3.0, respectively.  }
\end{figure}

      We found in the cases of the Sun and Procyon that the collisions with hydrogen atoms can provide a marked contribution to non-LTE effects for the C I lines (excluding the lines in the visual region which are insensitive to them). For instance, when calculating the collisional rates from Steenbock \& Holweger's (1984) formula with the scaling factor $\rm{k_{h}}=0.0$ and $\rm{k_{h}}=1.0$, we obtained for the C I lines of the IR multiplet No. 2 a difference about 0.15 dex in the C abundances for the Sun. Adopting the factor $\rm{k_{h}}= 0.3-0.6$ we could reach agreement between various lines in the derived C abundances (in particular, between lines in the visual and IR regions); so, we adopted $\rm{k_{h}}= 0.4$ in our further computations. It is important to note that the Sun and Procyon are dwarfs with $\log g$ $\sim$ 4; in the case of supergiants and bright giants with $\log g$~= 1.0-2.7 studied in the present work a role of collisions with hydrogen is significantly lowered. 

      The adopted carbon model atom allowed us to describe observed profiles of all C I lines by one carbon abundance; namely, we found $\log \epsilon$(C)~= 8.43 for the Sun and $\log \epsilon$(C)~= 8.40 for Procyon. It should be noted that our $\log \epsilon$(C) value for the Sun coincides with the solar value of Asplund et al. (2009). 

      Effective temperatures $T_{\rm eff}$ of these two stars are rather low, namely 5780~K for the Sun and 6590~K for Procyon (Korn et al. 2003). Our sample of programme stars includes objects up to the temperature $T_{\rm eff}$~= 8600~K; so, it is important to analyze departures from LTE in the whole $T_{\rm eff}$ range of interest. Basing on our computations, we show in Fig.2 the non-LTE correction for $\log \epsilon$(C) relatively to LTE as a function of $T_{\rm eff}$ for the surface gravities $\log g$~= 2.0 (upper panel) and $\log g$~= 3.0 (lower panel). One may see that in the $T_{\rm eff}$ range from 5000 to 8700~K the non-LTE corrections are negative; moreover, they tend to increase with increasing $T_{\rm eff}$. The corrections are rather small for lines in the visual region; even for hottest stars with $T_{\rm eff}$~= 8600~K they are less than 0.2 dex. However, in the case of IR lines with wavelengths from 9111 to 9658~\AA\ (multiplets No. 2, 3 and 9) the corrections are significant and can reach -1.0 dex for supergiants with $T_{\rm eff}$~= 7500-8700~K and $\log g$~= 2.0.

      What is a reason of such a difference in non-LTE effects between lines in the visual and IR regions? According to our non-LTE computations, lower levels $3s3P^{0}$ and $3s1P^{0}$ of the strong IR lines of multiplets No. 2, 3 and 9 show a large over-population at depths where these lines are formed ($\log \tau_{5000}$ $\sim$ -2 to -3), whereas their upper levels show relatively small under-population as compared with LTE. On the other hand, the C I lines in the visual region are formed in deeper layers ($\log \tau_{5000}$ $\sim$ -1 to -0.5), where collisions with electrons lead to decrease of non-LTE effects. As a result, populations of both upper and lower levels of these lines are close to the equilibrium populations; only slight under-population is seen as compared with LTE. 

      One may see from Fig.2 that the non-LTE corrections for $\log g = 2.0$ are greater than for $\log g = 3.0$. Moreover, non-LTE effects in C I lines depend on the adopted carbon abundance $\log \epsilon$(C), as well as the microturbulent parameter $V_t$ (note that in computations presented in Fig.2 we adopted the solar C abundance and $V_t$~= 2~km/s). Therefore, individual non-LTE computations are needed for each programme star.

\section{LIST OF PROGRAMME STARS AND THEIR EVOLUTIONARY STATUS}

\begin{table*}
 \centering
 \begin{minipage}{110mm}
 \caption{List of 36 programme stars with their parameters }
  \begin{tabular}{|c|c|c|c|c|c|c|c|c|c|}
  \hline
HR&HD&Name&Sp&$T_{\rm eff}$, K&$\log g$&$V_t$, &$d$, pc&$M/M_\odot$&$t$,
\\
&&&&&&&km/s&&$10^{6}$ yr\\
  \hline
292&6130&&F0 II&6880&2.05 &2.7&613&7.1&44\\
825&17378 &&A5 Ia&8570&1.18&10.8&2700&23.9&  7\\
1017&20902&$\alpha$ Per&F5 Ib&6350&1.90&5.3&156&7.3&41\\
1242&25291 &&F0 II&6815&1.87&3.2&629&8.3&32\\
1303&26630&$\mu$ Per&G0 Ib&5380&1.73&3.6&275&7.0&46\\
1603&31910&$\beta$ Cam&GI Ib-IIa&5300&1.79&4.8&265&6.5&53\\
1740&34578&19 Aur&A5 II&8300&2.10&4.3&637&8.8&29\\
1865&36673&$\alpha$ Lep&F0 Ib&6850&1.34&3.9&680&13.9&13\\
2000&38713&&G5 II&5000&2.45&2.1&224&3.9&184\\
2597&51330&&F2 Ib-II&6710&2.02&3.3&935&7.1&44\\
2693&54605&$\delta$ CMa&F8 Ia&5850&1.00&7.0&495&14.9&12\\
2786&57146&&G2 Ib&5260&1.90&3.2&386&5.9&67\\
2839&58585&&A8 I-II&7240&1.92&2.0&1330&8.6&30\\
2874&59612&&A5 Ib&8620&1.78&7.8&940&12.9&15\\
2933&61227&&F0 II&6690&2.02&2.7&790&7.0&45\\
3102&65228&11 Pup&F7 II&5690&2.17&3.7&161&5.1&92\\
3183&67456&&A5 II&8530&2.67&3.5&481&5.4&83\\
3291&70761&&F3 Ib&6600&1.25&3.9&2900&14.2&13\\
3459&74395&&G1 Ib&5370&2.08&3.5&236&5.2&89\\
5165&118605&83 Vir&G0 Ib-IIa&5430&2.37&2.5&253&4.2&152\\
6081&147084&$o$ Sco&A5 II&8370&2.12&2.8&269&8.7&29\\
6144&148743&&A7 Ib&7400&1.80&4.8&1330&10.0&23\\
6978&171635&45 Dra&F7 Ib&6000&1.70&4.6&649&8.2&33\\
7014&172594&&F2 Ib&6760&1.66&4.6&1000&10.0&22\\
7094&174464&&F2 Ib&6730&1.75&3.4&855&9.1&26\\
7264&178524&$\pi$ Sgr&F2 II&6590&2.21&3.2&156&5.9&67\\
7387&182835&32 Aql&F3 Ib&6700&1.43&4.4&880&12.5&15\\
7456&185018&&G0 Ib&5550&2.06&2.8&370&5.5&79\\
7542&187203&&F8 Ib-II&5750&2.15&4.2&376&5.3&86\\
7770&193370&35 Cyg&F5 Ib&6180&1.53&5.0&960&10.0&22\\
7796&194093&$\gamma$ Cyg&F8 Ib&5790&1.02&5.2&562&14.5&12\\
7823&194951&&F1 II&6760&1.92&4.2&1010&7.8&36\\
7834&195295&41 Cyg&F5 II&6570&2.32&3.6&235&5.3&85\\
7847&195593&44 Cyg&F5 Iab&6290&1.44&4.1&1040&11.2&18\\
7876&196379&&A9 II&7020&1.66&3.4&1740&10.6&20\\
8232&204867&$\beta$ Aqr&G0 Ib&5490&1.86&3.7&165&6.4&56\\
   \hline
   \end{tabular}
   \end{minipage}
\end{table*}

      Programme A-, F- and G-type stars are listed in Table 2, 36 objects in all. The following data are presented: HR an HD numbers of stars, their names (if they exist) and spectral classification. Next we provide the stellar parameters from Paper I, namely the effective temperature $T_{\rm eff}$, surface gravity $\log g$, microturbulent parameter $V_t$, distance \textit{d}, mass \textit{M} in solar masses M$_\odot$, and age \textit{t}. 

\begin{figure}
\begin{center}
\includegraphics[width=83mm]{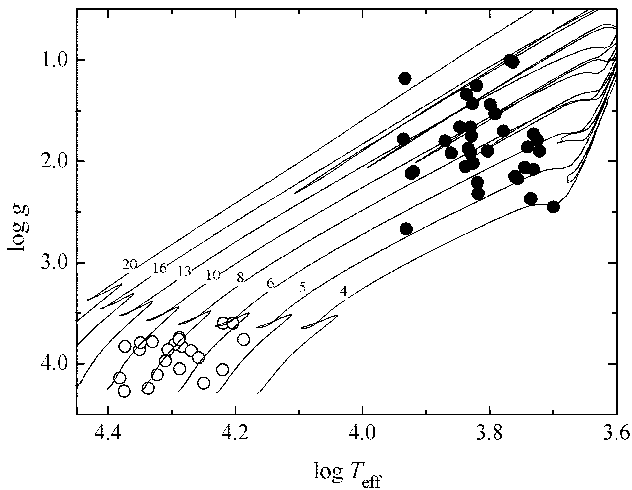}
\end{center}
\caption[]{Evolutionary tracks of Claret (2004) on the $T_{\rm eff}$~-- $\log g$ plane. Programme AFG supergiants and bright giants are shown by filled circles, B-type MS stars from Lyubimkov et al. (2013) are presented by open circles. }
\end{figure}

      Following Papers I, II and III, we show in  Fig.3 the positions of the programme stars on the $T_{\rm eff}$--$\log g$ diagram together with Claret's (2004) evolutionary tracks. Also in Fig.3, we show the 22 B-type MS stars, whose C, N and O abundances have been analyzed by Lyubimkov et al. (2013). Programme AFG stars are shown by filled circles and the B-type MS stars are by open circles. It should be noted that the same method for both types of the stars was applied for the $\log g$ determination: this method is based on van Leeuwen's (2007) stellar parallaxes. It was shown in Paper I that this method allows significant improvement in the accuracy of the $\log g$ determination in comparison with previous evaluations. 

      One may see from Fig.3 that the B-type MS stars and the AFG supergiants and bright giants are successive evolutionary phases of stars of the same masses. The B stars presented in Fig.3 have masses \textit{M} between 5 and 11 M$_\odot$, whereas masses of the programme AFG supergiants and bright giants spread mostly from 4 to 15 M$_\odot$ with the especially high value \textit{M}~= 23.9 M$_\odot$ for HR 825. 

      As noted in the Introduction, atmospheric C and N abundances may change during the MS phase because of the rotationally induced mixing, if a star is rather massive and its rotational velocity is rather high. Additional changes in C and N abundances occur during the AFG supergiant phase due to the first dredge-up (the FD phase). We discuss both these evolutionary phases below. 

      The well-known interesting features of evolutionary tracks during the FD phase are seen in Fig.3, namely the red-blue loops. Problem of the loops was considered in more details in Paper II. It is important that in 'the loop area' both the post-MS stars and the post-FD supergiants can be mixed. Separation of these two sorts of stars is possible only on a basis of observed differences in their chemical anomalies. 
      
\section{DETERMINATION OF THE CARBON ABUNDANCES}

      Model atmospheres for the stars were computed applying Kurucz's (1993) code ATLAS9 for the adopted $T_{\rm eff}$, $\log g$ and $V_t$ values. Using these model atmospheres and the microturbulent parameter $V_t$ from Table 2, we determined the carbon abundance $\log \epsilon$(C) for each star. In order to take into account the blending by lines of other chemical elements, we based our analysis on computations of synthetic spectra and their comparison with observed ones. The accuracy of the $\log \epsilon$(C) determination from an individual line is within 0.05-0.10 dex. The number of lines used varies from 6 to 26 and depends partially on the spectrum blending by telluric lines. 

      Computation of synthetic spectra and determination of abundances $\log \epsilon$(C) require adoption of a projected rotational velocity \textit{v} $\sin i$ for each star. It was noted in Paper III that when spectra of cool supergiants and bright giants are studied, it is difficult to separate correctly the contributions to a line profile of the projected rotational velocity \textit{v} $\sin i$ and the macroturbulent velocity $V_{mac}$. Two limiting cases were considered in Paper III: calculation of spectra with only \textit{v} $\sin i$ or only with $V_{mac}$; it was found that the derived Li abundances for these two cases are identical within 0.03 dex. So, following Paper III, we determine in the present work the \textit{v} $\sin i$  values with neglect of the $V_{mac}$ (therefore, the derived rotational velocities \textit{v} $\sin i$  are upper limits to the actual \textit{v} $\sin i$ ). 

      The \textit{v} $\sin i$ values determined from C I lines are presented in Table 3 together with the C abundances. One sees that the rotational velocities of the 36 programme stars vary from 4 to 35~km/s. For 17 of these stars the \textit{v} $\sin i$ values have been obtained in Paper III from Fe I lines in the vicinity of the Li I 6708~\AA\ line. Comparison of \textit{v} $\sin i$ for these common 17 stars from two works shows very good agreement; the difference is within $\pm$1.0~km/s. 

      We display in Fig.4, as an example of our technique, the fit of the computed non-LTE profiles of four C I lines (solid curves) to the observed profiles (dots) for two stars; note that these C I lines span the range of wavelengths from 5380 to 9112~\AA. It should be noted as well that there is a marked difference between effective temperatures of the stars: $T_{\rm eff}$~= 6350~K for HR 1017 and 8370~K for HR 6081. It is important that in both cases we obtained good agreement between the computed and observed profiles of all lines by using the one $\log \epsilon$(C) value for each star. One sees from Fig.4 that the LTE profiles calculated with the same $\log \epsilon$(C) values (dotted curves) are much shallower than the observed ones, especially for IR lines. 

\begin{figure}
\begin{center}
\includegraphics[width=83mm]{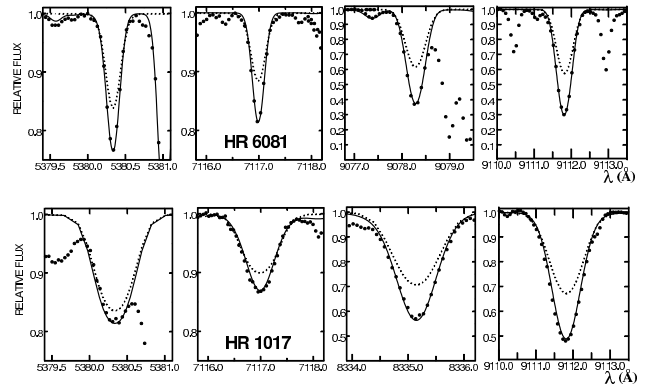}
\end{center}
\caption[]{Comparison of the observed and computed profiles of some C I lines for the stars HR 1017 and HR 6081. Solid curves~-- non-LTE, dotted curves~-- LTE. }
\end{figure}

\begin{figure}
\begin{center}
\includegraphics[width=83mm]{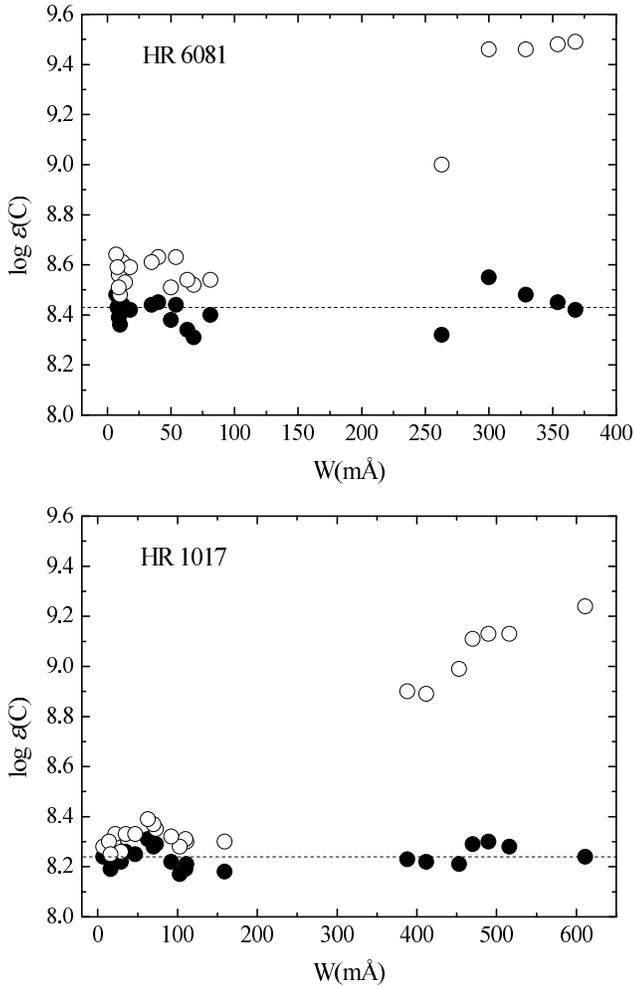}
\end{center}
\caption[]{The non-LTE and LTE carbon abundances (filled and open circles, respectively) for the stars HR 1017 and HR 6081 as a function of the equivalent width \textit{W} of C I lines. The dashed horizontal lines correspond to mean non-LTE abundances. }
\end{figure}

\begin{figure}
\begin{center}
\includegraphics[width=83mm]{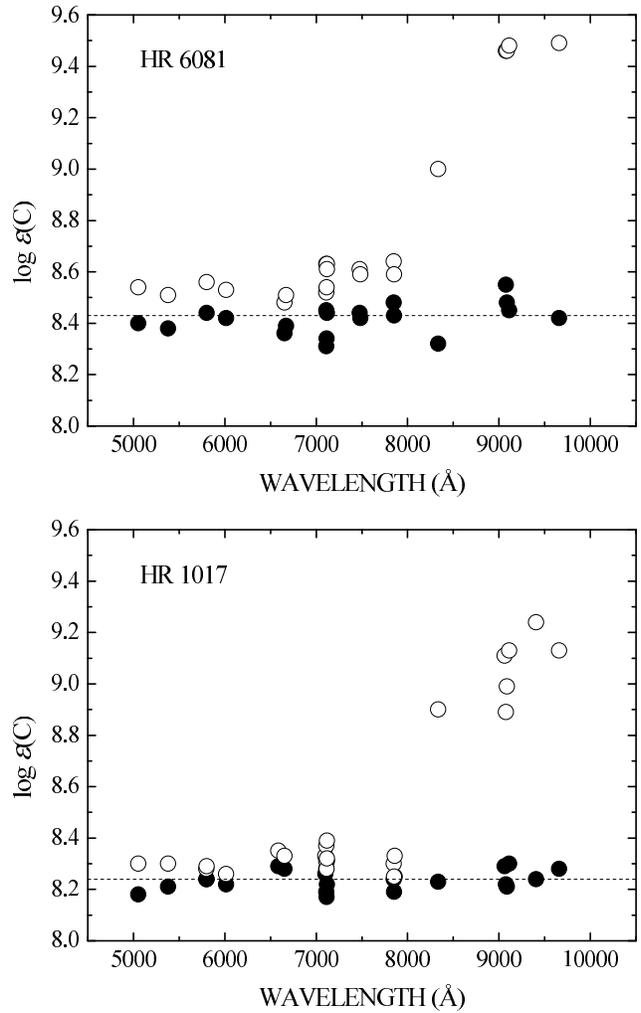}
\end{center}
\caption[]{The non-LTE and LTE carbon abundances for the stars HR 1017 and HR 6081 as a function of the wavelength of C I lines. The dashed horizontal lines correspond to mean non-LTE abundances.}
\end{figure}

      Comparison of the LTE and non-LTE carbon abundances for the same stars HR 1017 and HR 6081 is presented in Figs 5 and 6; individual $\log \epsilon$(C) values for all used C I lines are displayed there. In Fig.5 the abundances $\log \epsilon$(C) are shown as a function of the equivalent width \textit{W}. One may see that the non-LTE abundances (filled circles) lie very compactly around the mean values (dashed horizontal lines), whereas the LTE abundances (open circles) determined for the same microturbulent parameters $V_t$ reveal a steep trend with \textit{W}. It is important that all strongest lines in Fig.5 are situated in the IR region; so, according to Fig.6, the trend of the LTE abundances with the wavelength takes place, too.

      In Table 3 we present main quantitative results of our analysis. We show there for each star the number of C I lines used, the projected rotational velocity \textit{v} $\sin i$, the mean carbon abundance $\log \epsilon$(C) with corresponding error, and also the [\textit{C/H}], [\textit{Fe/H}] and [\textit{C/Fe}] values. The [\textit{Fe/H}] values (indicators of metallicity) are taken from Paper I. Two other values, [\textit{C/H}] and [\textit{C/Fe}] give the C abundances relative to the Sun, namely: [\textit{C/H}]~= $\log \epsilon$(C)~-- $\log \epsilon_{\odot}$(C) and [\textit{C/Fe}]~= $\log \epsilon$(C/Fe)~-- $\log \epsilon_{\odot}$(C/Fe), where $\log \epsilon$(C/Fe)~= $\log \epsilon$(C)~-- $\log \epsilon$(Fe).

\begin{table*}
 \centering
 \begin{minipage}{140mm}
 \caption{The derived rotational velocities and carbon abundances for programme stars, as well as previous data for their Fe and N abundances}
  \begin{tabular}{|c|c|c|c|c|c|c|c|c|c|}
  \hline
HR&Number of &$v \sin i$, &$\log \epsilon$(C)&[\textit{C/H}]&[\textit{Fe/H}]&[\textit{C/Fe}]&$\log \epsilon$(N)&[\textit{N/Fe}]&[\textit{N/C}]\\
& C I lines& km/s&&&&&&&\\
   \hline
292&15&14.0&8.14�0.12&-0.29&+0.05&-0.34&8.49�0.12&0.61&0.95\\
825&10&23.0&8.42�0.17&-0.01&-0.07&+0.06&7.90�0.10&0.14&0.08\\
1017&23&18.0&8.24�0.12&-0.19&-0.07&-0.12&8.41�0.10&0.65&0.77\\
1242&23&7.5&8.08�0.07&-0.35&-0.07&-0.28&8.34�0.10&0.58&0.86\\
1303&13&10.0&8.08�0.10&-0.35&-0.09&-0.26&--&--&--\\
1603&15&11.0&8.20�0.18&-0.23&-0.04&-0.19&--&--&--\\
1740&18&8.0&8.23�0.11&-0.20&-0.08&-0.12&8.12�0.07&0.37&0.49\\
1865&17&10.5&7.77�0.15&-0.66&+0.03&-0.69&8.73�0.09&0.87&1.63\\
2000&10&6.0&8.20�0.17&-0.23&-0.10&-0.13&--&--&--\\
2597&12&35.0&8.04�0.25&-0.39&-0.18&-0.21&8.15�0.29&0.50&0.71\\
2693&11&16.0&8.24�0.20&-0.19&+0.01&-0.20&8.45�0.16&0.61&0.81\\
2786&6&10.0&8.23�0.17&-0.20&+0.05&-0.25&--&--&--\\
2839&16&4.0&7.99�0.08&-0.44&-0.10&-0.34&8.39�0.14&0.66&1.00\\
2874&11&12.0&8.34�0.15&-0.09&+0.02&-0.11&8.24�0.10&0.39&0.50\\
2933&19&5.0&8.01�0.08&-0.42&-0.13&-0.29&8.40�0.17&0.70&0.99\\
3102&11&12.0&8.30�0.20&-0.13&+0.11&-0.24&8.35�0.25&0.41&0.65\\
3183&13&23.0&8.39�0.15&-0.04&+0.04&-0.08&8.12�0.04&0.25&0.33\\
3291&12&13.0&7.95�0.18&-0.48&-0.09&-0.39&8.65�0.15&0.91&1.30\\
3459&14&10.0&8.18�0.17&-0.25&+0.03&-0.28&--&--&--\\
5165&7&10.0&8.15�0.20&-0.28&-0.17&-0.11&--&--&--\\
6081&20&4.0&8.43�0.06&0.00&+0.03&-0.03&8.30�0.07&0.44&0.47\\
6144&14&15.0&7.89�0.14&-0.54&-0.11&-0.43&8.60�0.23&0.88&1.31\\
6978&14&10.0&8.04�0.09&-0.39&-0.09&-0.30&8.25�0.09&0.51&0.81\\
7014&26&11.0&8.23�0.15&-0.20&-0.07&-0.13&8.36�0.16&0.60&0.73\\
7094&16&17.0&8.08�0.15&-0.35&-0.18&-0.17&8.15�0.16&0.51&0.68\\
7264&11&25.0&8.08�0.20&-0.35&-0.17&-0.18&8.30�0.12&0.64&0.82\\
7387&22&9.0&8.00�0.15&-0.43&-0.03&-0.40&8.61�0.13&0.81&1.21\\
7456&16&9.0&8.05�0.12&-0.38&-0.16&-0.22&--&--&--\\
7542&11&25.0&8.30�0.20&-0.13&+0.17&-0.30&--&--&--\\
7770&18&4.5&8.10�0.15&-0.33&-0.22&-0.11&8.14�0.12&0.53&0.64\\
7796&16&9.0&8.03�0.10&-0.40&-0.04&-0.36&8.28�0.13&0.49&0.85\\
7823&19&19.0&8.07�0.15&-0.36&-0.13&-0.23&8.43�0.14&0.73&0.96\\
7834&19&9.0&8.22�0.15&-0.21&0.00&-0.21&8.20�0.11&0.37&0.58\\
7847&19&6.5&8.18�0.10&-0.25&-0.06&-0.19&8.34�0.14&0.57&0.76\\
7876&21&25.0&8.22�0.12&-0.21&-0.21&+0.00&7.75�0.13&0.13&0.13\\
8232&14&10.5&8.25�0.15&-0.18&+0.10&-0.28&--&--&--\\
   \hline
   \end{tabular}
 \end{minipage}
\end{table*}

      In three last columns of Table 3 we provide the nitrogen abundances $\log \epsilon$(N) and [\textit{N/Fe}] from Paper II; these values are known for 27 of 36 programme stars. The nitrogen-to-carbon ratio relative to the Sun is given as well, i.e. [\textit{N/C}]~= [\textit{N/Fe}]~-- [\textit{C/Fe}].  Note that use of the relative abundances [\textit{C/Fe}] and [\textit{N/Fe}] allows to take into account the difference in metallicities of the stars. 

      We adopted in Table 3 for the [\textit{C/Fe}] and [\textit{N/Fe}] evaluation the solar abundances from Asplund et al. (2009), namely $\log \epsilon_{\odot}$(C)~= 8.43$\pm$0.05, $\log \epsilon_{\odot}$(N)~= 7.83$\pm$0.05 and $\log \epsilon_{\odot}$(Fe)~= 7.50$\pm$0.04; therefore, $\log \epsilon_{\odot}$(C/Fe)~= 0.93. However, we shall use also for comparison the solar N and C abundance from Caffau et al. (2009, 2010), namely $\log \epsilon_{\odot}$(N)~= 7.86$\pm$0.12 and $\log \epsilon_{\odot}$(C)~= 8.50$\pm$0.06; the iron abundance is also set at $\log \epsilon_{\odot}$(Fe)~= 7.50. 

      It should be noted that both Asplund et al.'s (2009) solar abundances and Caffau et al.'s (2009, 2010) data have been obtained on the basis of three-dimensional (3D) hydrodynamical models of the solar atmosphere. It is interesting that our estimation of the solar carbon abundance determined above with the 1D-model of the solar atmosphere (Castelli \& Kurucz 2003) provided $\log \epsilon$(C)~= 8.43, a value equal to Asplund et al.'s value. 

      Strictly speaking, one should reference the derived C and N abundances to the C and N abundances in early B-type main sequence (MS) stars, the immediate progenitors of AFG supergiants and giants. It should be noted in this connection that contemporary data confirm a similarity of the C and N abundances for the unevolved B-type MS stars and the Sun. In particular, our recent analysis (Lyubimkov et al. 2013) of CNO abundances for a sample of early B stars gave mean values $\log \epsilon$(C)~= 8.33$\pm$0.11 and $\log \epsilon$(N)~= 7.78$\pm$0.09; so, a good agreement with the current solar nitrogen abundance is found, but carbon is kind of slightly deficient. Meantime, more severe selection of the sampled B stars, when only the relatively cool stars with effective temperatures $T_{\rm eff}$ between 15000 and 18100~K are considered, where a possible over-ionization of C II ions is imperceptible, led to the mean C abundance $\log \epsilon$(C)~= 8.46$\pm$0.09 (Lyubimkov, 1913), a value in a very good agreement with the above-mentioned solar C abundances. So, the C and N anomalies in the stars of our interest may really be measured with respect to the solar values. Therefore, the cited solar abundances of Asplund et al.'s (2009) and Caffau et al.'s (2009, 2010) are used further as reference values.  

      We analyze below the relation between the C and N abundances. It should be noted in this connection that errors in the derived $\log \epsilon$(C) and $\log \epsilon$(N) values are comparable. In particular, according to Table 3, the mean error is $\pm$0.14 dex for carbon (36 stars) and $\pm$0.13 dex for nitrogen (27 stars). 

      It is interesting to compare our results with the recent Luck's (2014) paper, where basic parameters and element abundances were determined for 451 FGK stars of luminosity classes I and II. A study of the galactic abundance gradient was one of the main goals of this paper. There are 23 common objects in our list and Luck's paper. For all these 23 stars the C abundances were obtained by Luck; only for two of them the N abundances were found. When comparing the [\textit{C/Fe}] values from two works, one sees that the differences in [\textit{C/Fe}] between Luck and our data vary mostly between 0.20 and -0.25 dex. Moreover, a trend with $T_{\rm eff}$ is seen, so for stars with $T_{\rm eff} < 5700$~K Luck's [\textit{C/Fe}] values are lower than ours by 0.2 dex on average. What are possible causes of such differences? It should be noted in this connection that the derived element abundances depend, first of all, on the accuracy of the basic parameters $T_{\rm eff}$ and $\log g$. Therefore, the methods used in these works for the $T_{\rm eff}$ and $\log g$ determination are of special interest. 

   1) Effective temperatures $T_{\rm eff}$ were determined in these two works by different methods. As a result, there is a trend in the $T_{\rm eff}$ differences: the higher $T_{\rm eff}$, the greater the differences. So, Luck's values show the systematic overestimation for $T_{\rm eff} > 5700$~K, which reaches 300--400~K for $T_{\rm eff}$ $\sim$ 6800~K. Luck's technique of the $T_{\rm eff}$ determination is based on Kovtyukh's (2007) method for stars with $T_{\rm eff} < 6800$~K (for a discussion of this method see in Paper I), whereas for stars with $T_{\rm eff} > 6800$~K an excitation analysis of Fe I and Fe II lines is used. Meanwhile, Fe I lines in spectra of F supergiants are rather sensitive to non-LTE effects (Boyarchuk, Lyubimkov and Sakhibullin, 1985); so, ignoring these effects can lead to underestimation of Fe abundances and, as a result, to systematic errors in the $T_{\rm eff}$ evaluation. 

   2) Surface gravities $\log g$ were derived by different methods, too. The differences in $\log g$ values are mostly within $\pm$0.3 dex; note that they show no trend with $T_{\rm eff}$. Luck's $\log g$ values were based on the ionization balance for Fe I and Fe II lines. As mentioned above, Fe I lines in spectra of yellow supergiants are sensitive to non-LTE effects; ignoring these effects can lead to errors in the $\log g$ evaluation. Our method of the $\log g$ determination is based on use of stellar parallaxes (Paper I); undoubtedly, this method is most accurate for bright and nearby stars, like ours.  

   3) It should be noted that the [\textit{Fe/H}] values (index of metallicity) for 23 common stars do not show any systematic differences in two works. Our mean value for these stars is [\textit{Fe/H}]~= -0.05$\pm$0.10, whereas Luck's value is [\textit{Fe/H}]~= 0.00$\pm$0.10.             

   4) It is important to note as well that Luck's carbon abundances are based on the LTE approach, whereas our C abundances are inferred from a non-LTE analysis. Luck's abundances were derived from C I lines at 5052, 5380 and 7115~\AA, as well as C$_{2}$ lines at 5135~\AA. As mentioned above (see Fig.2), the non-LTE corrections are rather small for C I lines in the visual region (like lines at 5052 and 5380~\AA), but for the line 7115~\AA\ in near IR region the correction can reach -0.15 dex at $T_{\rm eff} \sim 7500$~K. 

      Summarizing, we may conclude that some discrepancy between our and Luck's [\textit{C/Fe}] values can be partially due to differences in the adopted basic parameters $T_{\rm eff}$ and $\log g$. This example shows once again that these parameters should be determined with a high accuracy, when fine evolutionary effects in observed abundances are studied. We believe that our results have the high accuracy that is confirmed by their good agreement with theoretical predictions (see below). 

\section{CARBON ABUNDANCE AS A FUNCTION OF PARAMETERS $T\lowercase{_{\rm eff}}$, $\log\lowercase{g}$ AND $M/M_{\odot}$ }

      As the initial step in interpretation of the derived carbon abundances $\log \epsilon$(C), we consider relations between $\log \epsilon$(C), on the one hand, and two basic stellar parameters, namely the effective temperature $T_{\rm eff}$ and surface gravity $\log g$, on the other hand. These relations are presented in Fig.7, where we show the C abundances for 36 programme stars (open circles) with error bars, as well as the solar C abundances (dashed horizontal lines) taken from two sources, namely A: $\log \epsilon_{\odot}$(C)~= 8.43 (Asplund et al. 2009) and C: $\log \epsilon_{\odot}$(C)~= 8.50 (Caffau et al. 2010). An evident important conclusion follows from Fig.7 at once: across the $T_{\rm eff}$ interval from 5000~K to 8620~K (upper panel) and the $\log g$ interval from 1.0 to 2.7 dex (lower panel) most programme AFG supergiants and bright giants show a marked carbon underabundance as compared with the Sun; moreover, there is no star with the $\log \epsilon$(C) value greater than the solar one. Therefore, we confirmed the carbon deficiency in atmospheres of this type stars as their general property. According to Fig.7, the C deficiency can vary from zero down to -0.6 dex. 

\begin{figure}
\begin{center}
\includegraphics[width=83mm]{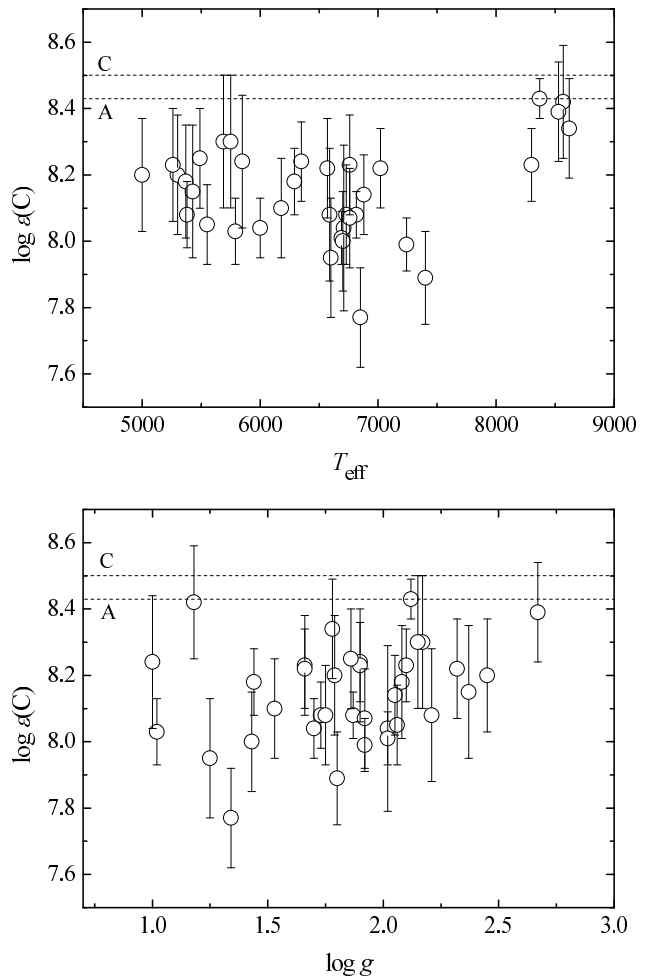}
\end{center}
\caption[]{The derived carbon abundance for 36 programme stars as a function of $T_{\rm eff}$ and $\log g$. Two solar C abundances are shown by dashed lines, namely $\log \epsilon_{\odot}$(C)~= 8.43~-- A (Asplund et al. 2009) and $\log \epsilon_{\odot}$(C)~= 8.50~-- C (Caffau et al. 2010). }
\end{figure}

      No appreciable trend with $T_{\rm eff}$ is seen for stars with $T_{\rm eff} < 8000$~K in Fig.7 (upper panel). However, the five hottest A-type stars with $T_{\rm eff}$~= 8300--8620~K, namely HR 825, 1740, 2874, 3183 and 6081, form there a compact group with a small C deficiency (relative to the Sun). We shall show below that all these stars are likely the post-MS objects, so an absence of the substantial C deficiency (as well as a presence of the moderate N excess) for them is explainable from the theoretical point of view. The conclusion that these A stars did not reach the FD phase, agrees with their location on the $T_{\rm eff}$--$\log g$ diagram (Fig.3). 

      As far as the dependence on $\log g$ is concerned, one may see from Fig.7 (lower panel) that the C deficiency tends to show the greater scatter for stars with lower $\log g$. As found in Paper I, the boundary between giants and supergiants lies at $\log g$ $\sim$ 2.0; therefore, the stars with $\log g < 2.0$ are mostly supergiants. Moreover, the stars with $\log g < 2.0$ are rather massive (see Fig.3). Since there are both post-MS and post-FD objects among such supergiants, a large scatter in the C abundances for stars with $\log g < 2.0$ is explainable: the C deficiency expected from the post-FD phase is augmented by C deficiency produced in MS objects. 

\begin{figure}
\begin{center}
\includegraphics[width=83mm]{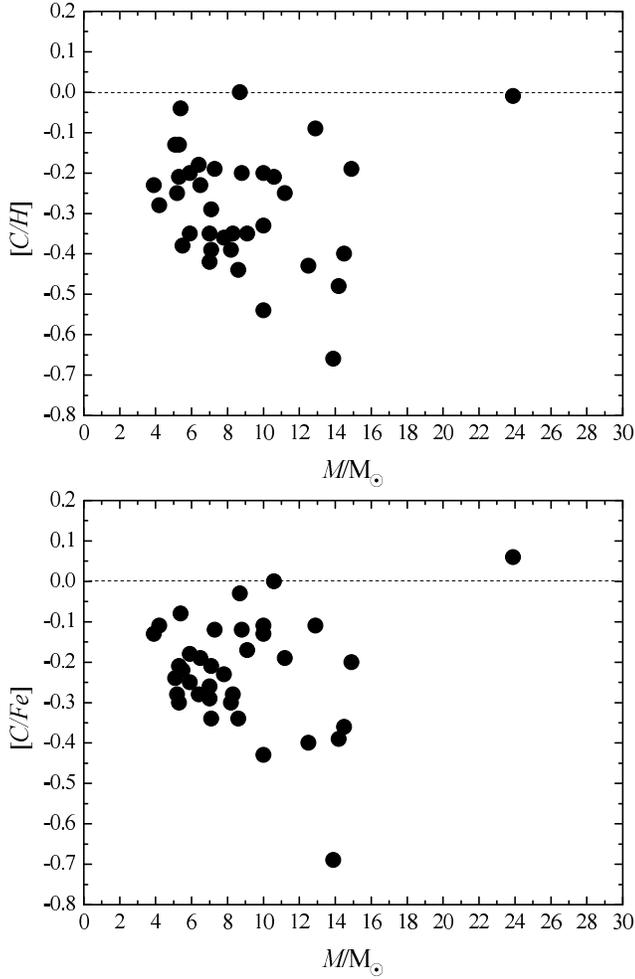}
\end{center}
\caption[]{The values [\textit{C/H}] and [\textit{C/Fe}] as a function of the mass $M/M_\odot$. Zero dashed line corresponds to Asplund et al.'s (2009) solar C and Fe abundances. }
\end{figure}

      Stellar mass \textit{M} is another basic parameter of great interest, as evolution of stars depends strongly on $M$. We show in Fig.8 relations between the relative carbon abundances [\textit{C/H}] and [\textit{C/Fe}] (presented in Table 3), on the one hand, and the mass $M/M_\odot$, on the other hand. The value [\textit{C/Fe}] takes into account a difference in the star's metallicity and is an appropriate quantity provided that C scales strictly with Fe in a star's initial composition. The masses \textit{M} of programme stars span the range from 4 to 24 M$_\odot$. There is a suggestion from Fig.8 that the scatter in carbon abundances tends to be greater for stars with greater masses. In fact, for stars with $M/M_\odot < 10$ the [\textit{C/Fe}] values vary from -0.03 to -0.34, whereas for stars with $M/M_\odot \geq 10$ these values vary from 0.06 to -0.69. Therefore, the scatter is about 0.3 dex for $M/M_\odot < 10$ and 0.75 dex for $M/M_\odot \geq 10$. 

\begin{figure}
\begin{center}
\includegraphics[width=83mm]{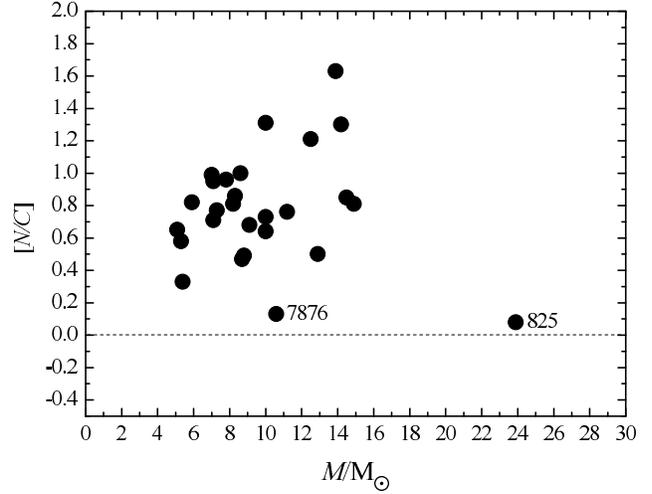}
\end{center}
\caption[]{The N/C ratio as a function of the mass $M/M_\odot$. The stars HR 825 and HR 7876 are marked, whose ratios are very close to the solar value (zero dashed line). }
\end{figure}

      According to the theory, the N/C ratio can be more sensitive to stellar evolution than the C or N abundances themselves, so the N/C dependence on the mass $M/M_\odot$ is of great interest. We show in Fig.9 the [\textit{N/C}] values from Table 3 as a function of $M/M_\odot$. If obvious 'outliers' HR 825 and HR 7876 are excluded, all stars occupy the [\textit{N/C}] region from 0.3 to 1.6. Again, like Fig.8, the scatter is greater for stars with $M/M_\odot \geq 10$ than for stars with $M/M_\odot < 10$. This may mean that, in agreement with the theory, the ratio [\textit{N/C}] for the more massive stars is more sensitive to the initial rotational velocity (see below).

\section{ANTICORRELATION BETWEEN THE CARBON AND NITROGEN ABUNDANCES}

      According to the theory as well as to preceding observational analyses, a N vs. C anticorrelation is expected for the AFG supergiant, i.e. for stars with masses \textit{M} between 4 and 25 M (the objects of our interest), when they are passing through the FD-phase. We show in Fig.10 a relation between $\log \epsilon$(N) and $\log \epsilon$(C) values for 27 stars of our list, for which both C and N abundances are available. The solar C and N abundances from two sources are shown for comparison (open circles), namely A: Asplund et al. (2009) and C: Caffau et al. (2009, 2010). 

\begin{figure}
\begin{center}
\includegraphics[width=83mm]{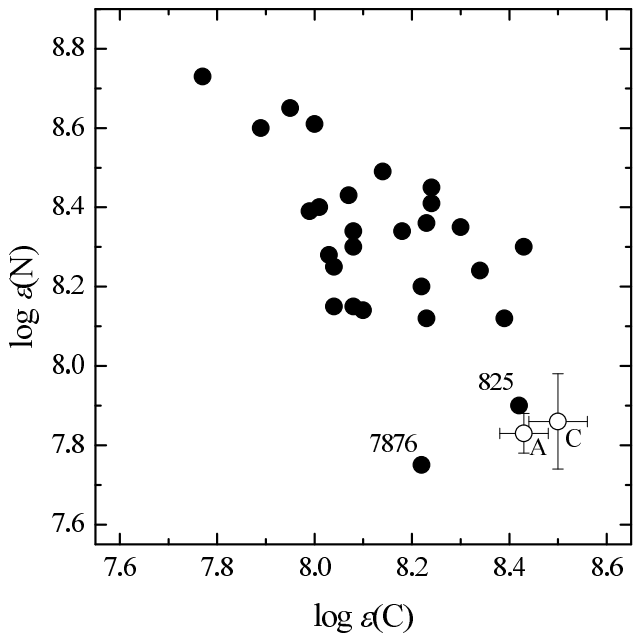}
\end{center}
\caption[]{Anti-correlation between the N and C abundances. Two open circles correspond to the solar N and C abundances from A~= Asplund et al. (2009) and C~= Caffau et al. (2009, 2010). The stars HR 825 and HR 7876 with the especially low N abundances are marked. }
\end{figure}

      The anticorrelation between the N and C abundances is clearly seen in Fig.10. Two stars with the lowest ($\sim$ solar) nitrogen abundance, HR 825 and HR 7876, are marked in Fig.10. The A5-supergiant HR 825, the most massive star in our list ($M \approx 24~\rm M_\odot$), has actually the solar C and N abundances, whereas the A9-bright giant HR 7876 shows an apparent C underabundance. However, according to Fig.11, when taking into consideration differences in the star's metallicity and pass on the relative abundances [\textit{C/Fe}] and [\textit{N/Fe}], we obtain good agreement with solar data for HR 7876, too. 

\begin{figure}
\begin{center}
\includegraphics[width=83mm]{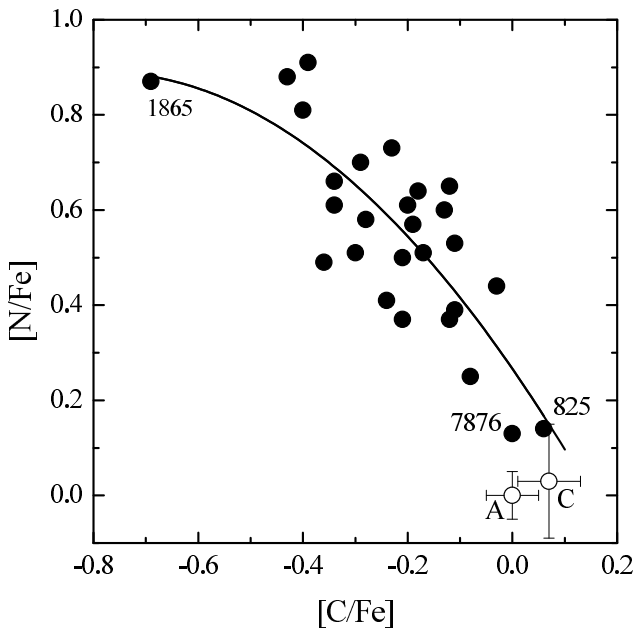}
\end{center}
\caption[]{Anti-correlation between the [\textit{N/Fe}] and [\textit{C/Fe}] values. These values are calculated relatively to Asplund et al.'s (2009) solar N and C abundances, so the open circle A has zero coordinates. The open circle C corresponds to Caffau et al.'s (2009, 2010) abundances and has coordinates [\textit{C/Fe}]~= 0.07 and [\textit{N/Fe}]~= 0.03. Solid curve is drawn by the least square method. }
\end{figure}

       As far as the solar C and N abundances in Fig.11 are concerned, i.e. open circles A and C, the first of them (Asplund et al. 2009) has zero coordinates, whereas the second one (Caffau et al. 2009, 2010) has coordinates [\textit{C/Fe}]~= 0.07 and [\textit{N/Fe}]~= 0.03. We approximated the observed N vs. C relation in Fig.11 by the two-order polynomial. One sees that this approximation (solid curve) corresponds better to Caffau et al.'s (2009, 2010) solar abundances. 

      The stars HR 825 (A5 Ia) and HR 7876 (A9 II), whose [\textit{C/Fe}] and [\textit{N/Fe}] values are very close to zero, are of special interest. It was noted in Paper II that our parameters $T_{\rm eff}$ and $\log g$ for these stars are in good agreement with previous determinations of other authors. We suppose that their C and N abundances are close to their initial values; in other words, they did not change markedly during star's evolution, up to their present supergiants phase. We show below that such a scenario is quite understandable from the viewpoint of current evolutionary theory. 

      One remark should be made in connection with the C vs. N anticorrelation presented in Fig.11. During the CNO-cycle in a star's interior the N nuclides are synthesized both in the CN-cycle and in the ON-cycle. Since the latter is operating deeper in the star then the CN-cycle, one may suppose that the observed N enrichment in atmospheres of the stars results mainly from the CN-cycle. In this case it would be expected that the sum of C and N abundances is constant. However, the loci C+N~= const in Fig.11 (beginning from the initial points A and C) lie markedly lower then the observed relation, e.g. by about 0.2 dex for [\textit{C/Fe}]~= -0.4 to -0.7. In other words, the N enrichment is underestimated. So, some conversion of O to N should be taken into consideration in addition to the C to N conversion. Therefore, it would be more correct, if the sum C+N+O is conserved. We have not yet measured the O abundances and are, therefore, unable to apply the sum rule, but we show in the next section that modern theoretical models calculated with the rotationally induced mixing give a very good explanation of the observed C vs. N anticorrelation presented in Fig.11.

\section{COMPARISON WITH THEORETICAL PREDICTIONS}

      One sees that the carbon underabundance (like the nitrogen overabundance, see Paper II) is a general property of the AFG supergiants and bright giants. When comparing the observed C and N anomalies with predictions of the theory, we consider computations for two evolutionary phases: (i) directly after the termination of the MS phase (post-MS phase); (ii) near the end of the first dredge-up (post-FD phase). The results for both the rotating and non-rotating models for these phases are discussed below.  

\begin{figure}
\begin{center}
\includegraphics[width=83mm]{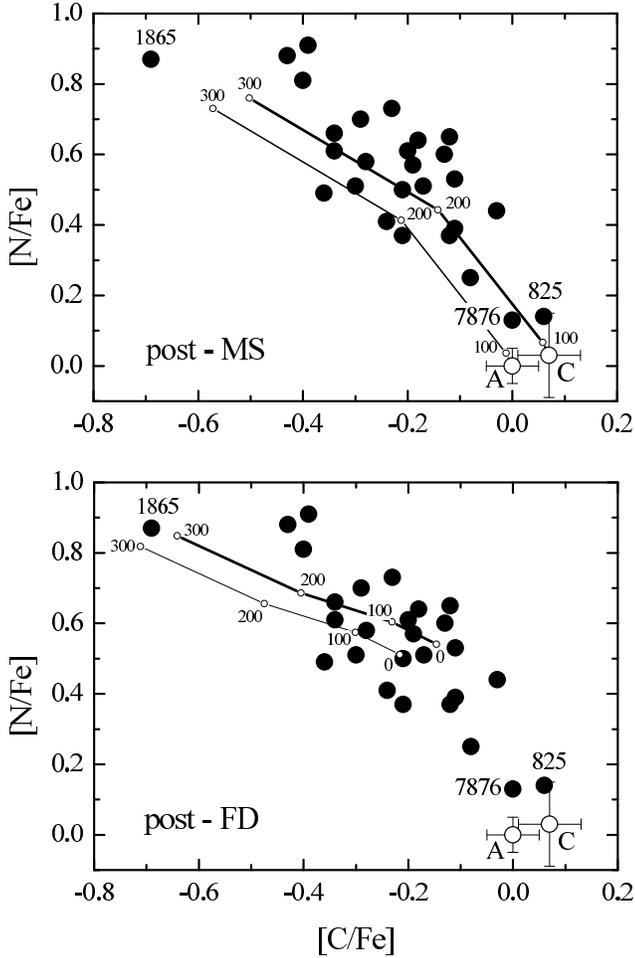}
\end{center}
\caption[]{Comparison of the observed [\textit{N/Fe}] vs. [\textit{C/Fe}] relation with theoretical predictions of Heger \& Langer (2000) for the 12 M$_\odot$ model. Like Fig.11, open circles A and C correspond to the solar N and C abundances from Asplund et al. (2009) and Caffau et al. (2009, 2010), respectively. Solid broken lines on the upper panel correspond to the predictions for the post-MS phase, on the lower panel~- for the post-FD phase; the thin and thick lines correspond to the initial points A and C, respectively. Initial rotational velocities 0, 100, 200 and 300~km/s are marked nearby the corresponding nodal points. Positions of the stars HR 825, 7876 and 1865 are marked. }
\end{figure}

      It is interesting to compare the observed anti-correlation between the N and C abundances shown in Fig.11 with the predicted relations for phases (i) and (ii). Such a comparison is presented in Fig.12; apart from the programme stars (filled circles), we show there by solid broken lines computations of Heger \& Langer (2000) for a 12 M$_\odot$ model. These computations are performed with the rotationally-induced mixing in the MS stage; we show in Fig.12 their results for the initial rotational velocities $V_0$~= 0, 100, 200 and 300~km/s (these values are marked in Fig.12 nearby the corresponding nodal points). Heger \& Langer's predictions for the post-MS phase and post-FD phase are presented separately on the upper and lower panels, respectively. It should be noted that Heger \& Langer calculated also models with masses \textit{M}~= 15 and 20 M$_\odot$, but differences between the relations [\textit{N/Fe}]--[\textit{C/Fe}] for various masses are rather small.  

      Thick broken lines in Fig.12 correspond to the initial point C (Caffau et al.'s solar abundances); thin lines correspond to the initial point A (Asplund et al.'s solar abundances). One may see that in the case C an agreement between the empirical and theoretical [\textit{N/Fe}]--[\textit{C/Fe}] relations seems better than in the case A for both post-MS and post-FD phase. 

      One sees from Fig.12 that Heger \& Langer's computations reproduce very well the observed [\textit{N/Fe}]--[\textit{C/Fe}] anti-correlation, especially in the case C. Some important conclusions follow from Fig.12. First of all, this anti-correlation reflects mostly a dependence of the C and N anomalies on the initial rotational velocity $V_0$: on average the higher the velocity $V_0$, the greater the anomalies. Therefore, Fig.12 confirms directly that the rotationally induced mixing in massive MS stars does really exist.  

      Next, Fig.12 shows that a cluster of the stars with [\textit{C/Fe}] between -0.1 and -0.4 and with [\textit{N/Fe}] between 0.3 and 0.7 may be identified as either post-MS objects with $V_0 \sim$ 200-250~km/s (upper panel) or post-FD objects with $V_0 \sim$ 0-150~km/s (lower panel). So, there is some ambiguity in evolutionary status of these stars. 

      The unusual stars HR 825 and HR 7876 without apparent [\textit{C/Fe}] and [\textit{N/Fe}] anomalies can be interpreted more definitely: they are post-MS objects with the small initial rotational velocities $V_0 \sim$ 100~km/s or less. Note as well that the star HR 1865, which shows the extreme C deficiency (marked in Figs 11 and 12) and seems a possible 'outlier' in Fig.11, corresponds very well to the general predicted relation (Fig.12, lower panel). This star with the greatest N/C ratio in Table 3 ([\textit{N/C}]~= 1.63) is very likely to be a post-FD object with $V_0 \sim$ 300~km/s. 

      The relation [\textit{N/C}] vs. $M/M_\odot$ presented in Fig.9 is also of great interest from the theoretical point of view. In Fig.13 we show both the observed [\textit{N/C}] values (filled circles) and the results of computations (broken lines) from two works, namely Heger \& Langer (2000) for models with masses M~= 12, 15 and 20 M$_\odot$ and Ekstr\"{o}m et al. (2012) for models over a much wider mass range. The computations for the post-MS phase (upper panel) and the post-FD phase (lower panel) are displayed. It should be noted that in these two works the C and N abundances are given as mass fractions; we converted these abundances into our scale (by number of atoms). 

\begin{figure}
\begin{center}
\includegraphics[width=83mm]{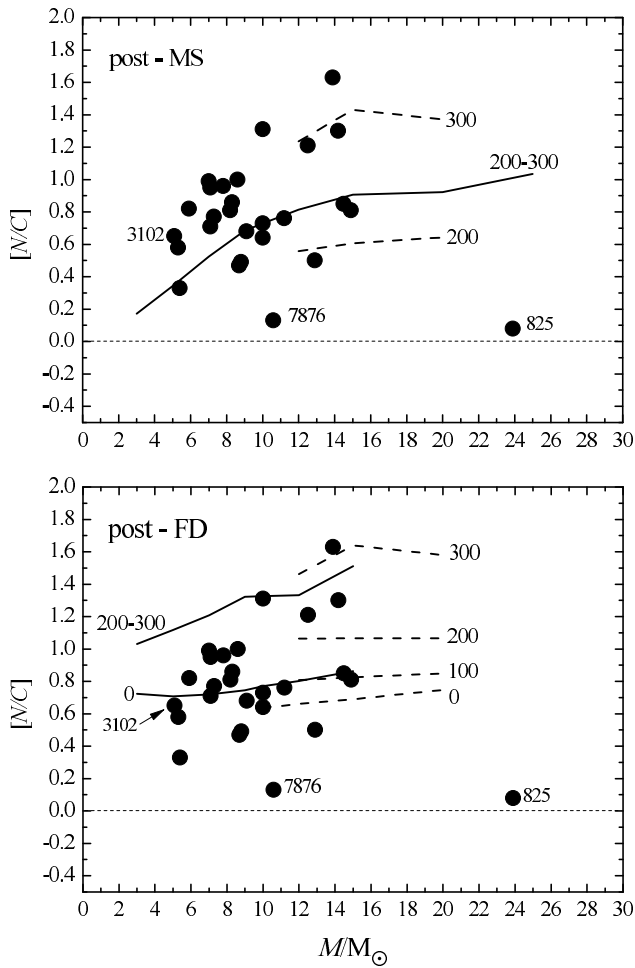}
\end{center}
\caption[]{Comparison of the observed [\textit{N/C}] vs. $M/M_\odot$ relation with theoretical predictions from two works, namely Heger \& Langer (2000) and Ekstr\"{o}m et al. (2012); dashed and solid broken lines correspond to the first work and the second one, respectively. Upper panel shows computations for the post-MS phase, lower panel presents the post-FD phase. Initial rotational velocities from 0 to 300~km/s are marked nearby the corresponding lines. Positions of the stars HR 825, 7876 and 3102 are marked. }
\end{figure}

      In the case of the post-MS phase (upper panel of Fig.13) Ekstr\"{o}m et al.'s results for models with \textit{M} from 3 to 25 M$_\odot$ are shown by one solid line with the initial rotational velocity $V_0 \approx$ 200--300~km/s; note that the adopted $V_0$ value in this work depends on the mass \textit{M} and equals 195~km/s for M~= 3~M$_\odot$ and 295~km/s for \textit{M}~= 25~M$_\odot$. When $V_0~= 0$~km/s (no rotation), theory predicts no changes in the surface [\textit{N/C}] values after the MS phase (zero dashed line in Fig.13). Results of Heger \& Langer for models with \textit{M}~= 12, 15 and 20~M$_\odot$ are presented there by two dashed broken lines with $V_0~= 200$ and 300~km/s; their results for $V_0~= 100$~km/s are not shown as they are very close to the case $V_0~= 0$~km/s (zero dashed line). One may see that there is good agreement between theoretical relations obtained from the rotational models with $V_0$~= 200--300~km/s, on the one hand, and the observed [\textit{N/C}] values, on the other hand. The stars HR 825 and HR 7876 are 'outliers' again; their low [\textit{N/C}] values confirm that they can be post-MS objects with $V_0$ $\sim$ 100~km/s or less. 

      In the case of the post-FD phase (lower panel of Fig.13) Ekstr\"{o}m et al.'s results are presented by two solid broken lines with $V_0~= 0$ and 200-300~km/s. Their [\textit{N/C}] values for \textit{M}~= 20 and 25 M$_\odot$ are not shown because for $V_0$~= 200--300~km/s they are much greater than the observed values ([\textit{N/C}]~= 2.5--2.6). Heger \& Langer's computations for models with \textit{M}~= 12, 15 and 20 M$_\odot$ are displayed by four dashed broken lines corresponding to the initial velocities $V_0$~= 0, 100, 200, and 300~km/s. It is interesting to note that even for $V_0$~= 0~km/s (non-rotating models) the computed [\textit{N/C}] values in the post-FD phase are significantly greater than zero. One may see that, from the viewpoint of the derived [\textit{N/C}] values, most of our programme stars with masses \textit{M} between 4 and 15 M$_\odot$ are the post-FD objects with the initial rotational velocities $V_0$ from 0 to 300~km/s. The stars HR 825 and HR 7876 are obvious 'outliers': their low [\textit{N/C}] are much lower than theoretical predictions even for $V_0$~= 0~km/s. These two stars can be only post-MS objects (see above). Four more stars with [\textit{N/C}]~= 0.3--0.5 (namely HR 1740, 2874, 3183 and 6081) are likely not post-FD objects; probably, they are post-MS stars with $V_0$ $\sim$ 200~km/s (see upper panel of Fig.13). 

      Thus, the comparison of our results with theoretical predictions reveals that the obtained C and N abundances for A-, F-, and G-type supergiants and bright giants are typical for the post-MS and/or post-FD stars with the initial rotational velocities $V_0$ from 0 to 300~km/s. In particular, the [\textit{N/C}] vs. $M/M_\odot$ relation, a very sensitive indicator of evolution, shows that the bulk of programme stars are either post-MS objects with $V_0$~= 200-300~km/s or post-FD objects with $V_0$~= 0-300~km/s. It is shown with confidence that the stars HR 825 and HR 7876 are the post-MS objects with $V_0$ $\sim$ 100~km/s or less; four more stars (HR 1740, 2874, 3183 and 6081) are likely post-MS objects with $V_0$ $\sim$ 200~km/s. Therefore, only 6 of 36 programme stars (i.e. 17 per cent) can be classified with some confidence as post-MS objects; nevertheless, the actual number of the stars in this phase can be greater. 

      The initial rotational velocity $V_0$ is an important parameter, on which the predicted C and N anomalies, as well as the anomalies of other light elements (e.g., He, Li and B) are strongly dependent. What actual $V_0$ values can be expected for stars with masses \textit{M} from 5 to 24 M$_\odot$? This question was discussed in details in Papers II and III. Basing on recent analyses of rotation velocities for B-type MS stars, the progenitors of supergiants and bright giants in question, one may conclude that the initial velocities $V_0$~= 100--300~km/s are quite common for such stars. Moreover, the fraction of stars with very low velocities (down to $V_0$ $\sim$ 0~km/s) is substantial, too. Therefore, the actual $V_0$ values correspond well to the $V_0$ values discussed above. 

      It is important to note that the $V_0$ value is the velocity at the beginning of the MS phase. When a star terminates this evolutionary phase and passes into the AFG supergiant phase, its rotational velocity decreases significantly. In particular, according to Table 3, our programme supergiants and bright giants have the projected rotational velocities $v \sin i$ from 4 to 35~km/s and for most of them $v \sin i < 20$~km/s. 

\section{CARBON AND NITROGEN ANOMALIES VS. LITHIUM DEPLETION}

      It is known that the lithium abundance $\log \epsilon$(Li) is a much more sensitive indicator of stellar evolution than the C and N abundances, so it would be interesting to compare the derived C and N abundances with known Li abundances. In Paper III we determined the $\log \epsilon$(Li) values for 55 F and G supergiants and bright giants; 17 of them are considered in the present work. Only for one of these 17 stars, HR 3102, was the lithium abundance derived with confidence: $\log \epsilon$(Li)~= 2.33 $\pm$ 0.19; the rest of the stars did not show the Li I 6708 line in their spectra, so it was possible to determine for them only an upper limit for $\log \epsilon$(Li). It was mentioned in Paper III that stars with $\log \epsilon$(Li) $\geq$ 2.0 are named the Li-rich stars, so the star HR 3102 belongs just to this type. The following important fact was noted in Paper III: all Li-rich giants and supergiants have masses  M $< 6 \rm~M_\odot$. 

      One sees from Table 3 that the bright giant HR 3102 (F7 II) with the mass \textit{M}~= 5.1 M$_\odot$ has the carbon deficiency [\textit{C/Fe}]~= -0.24, nitrogen excess [\textit{N/Fe}]~= 0.41 and, as a result, the rather high ratio [\textit{N/C}]~= 0.65. The position of this star in Fig.13 (marked there) shows that it can be, in principle, a post-MS object with $V_0 \sim 200$~km/s or a post-FD object with $V_0 \sim 0$~km/s. Both these possibilities for HR 3102 have been already discussed in Paper III. It has been concluded that the high Li abundance in the star's atmosphere, on the one hand, and simultaneously the enhanced N abundance (and now the lowered C abundance), on the other hand, cannot be explained by current theory. Therefore, it appears that the Li-rich giant HR 3102 may have freshly-produced lithium at the surface. 

      As far as the remaining 16 stars are concerned, for which only the $\log \epsilon$(Li) upper limits are known, it has been noted in Paper III that the absence of detectable lithium for the majority of such stars is explainable from the viewpoint of current theory. It is important that, according to our data, for all these 16 stars the C deficiency is found, namely [\textit{C/Fe}]~= -0.11 to -0.36. For about a half of them the N abundance is known from Paper II; the N excess takes place always, namely [\textit{N/Fe}]~= 0.37 to 0.61, so, the ratio [\textit{N/C}] is significantly enhanced. The observed C and N anomalies confirm that these 16 stars passed through the mixing during the MS and/or FD phase; therefore the strong depletion of lithium in their atmospheres is quite explainable. 
      
\section{CONCLUSIONS}

      Thus, our non-LTE analysis of the carbon abundance in 36 Galactic A-, F- and G-type supergiants and bright giants (luminosity classes I and II) led to the following conclusions. 

      First important result is that the most of programme stars show the carbon deficiency [\textit{C/Fe}] between -0.1 to -0.5 dex, with a minimum at -0.7 dex. So, the carbon underabundance as a general property of such stars is confirmed. 

      When comparing the derived carbon deficiency with the nitrogen excess determined for the same stars in Paper II, we obtain a pronounced C vs. N anti-correlation, as expected when CNO-cycled products are mixed into the atmosphere. 

      We found that the ratio [\textit{N/C}], a rather sensitive indicator of stellar evolution, spans mostly the range from 0.3 to 1.7 dex. Both these enhanced [\textit{N/C}] values and the C and N anomalous abundances themselves are an obvious evidence of the presence on a star's surface of mixed material from stellar interiors. 

      Our data on the surface C and N abundances in AFG supergiants and bright giants can be explained within the framework of stellar models with rotational mixing during the MS phase and the FD phase. Comparison with theoretical predictions shows that the observed N vs. C anti-correlation reflects a dependence of the C and N anomalies on the initial rotational velocity $V_0$: on average the higher $V_0$ the greater the anomalies. This fact confirms directly an existence of the rotational mixing. 

      The comparison with theoretical predictions shows that the bulk of programme stars are either post-MS objects with the initial rotational velocities $V_0$~= 200--300~km/s or post-FD objects with $V_0$~= 0--300~km/s. In particular, a cluster of the stars with [C/Fe] between -0.1 and -0.4 and with [\textit{N/Fe}] between 0.3 and 0.7 may be identified as either post-MS objects with $V_0$ $\sim$ 200--250~km/s or post-FD objects with $V_0$ $\sim$ 0-150~km/s. 

      Evolutionary status of the unusual stars HR 825 and HR 7876 with the almost unchanged C and N abundances is more definite: it is shown with confidence that they are the post-MS objects with $V_0$ $\sim$ 100~km/s or less. Four more stars with the slight C and N anomalies (HR 1740, 2874, 3183 and 6081) are likely the post-MS objects with $V_0$ $\sim$ 200~km/s. The star HR 1865 with large C and N anomalies and the greatest N/C ratio ([\textit{N/C}]~= 1.63) is very likely the post-FD object with $V_0$ $\sim$ 300~km/s.  

      Lithium, one more chemical element that is very sensitive to stellar evolution, is not visible in spectra of the majority of the AFG stars in question. Since the majority of these stars, according to our data, reveal the N excess and C deficiency, i.e. they passed through the deep mixing in the MS and/or FD phase, an absence of detectable lithium in their atmospheres is quite explainable from the theoretical point of view.     

\section*{Acknowledgments}

DLL acknowledges with thanks the support of the Robert A. Welch Foundation of Houston, Texas through grant F-634. SAK was supported by the SCOPES grant No. IZ73Z0-152485.

\label{lastpage}

\begin{thebibliography}{}
\bibitem{}
Allen C.W. 1976, Astrophysical Quantities, London: Athlone (3rd edition). 
\bibitem{}
Andrievsky S.M., Kovtyukh V.V., Korotin S.A., Spite M., Spite F. 2001, A\&A, 367, 605. 
\bibitem{}
Asplund M., Grevesse N., Sauval A.J., Scott P. 2009, ARA\&A, 47, 481. 
\bibitem{}
Bagnulo S., Jehin E., Ledoux C., Cabanac R., Melo C., Gilmozzi, R. 2003, Messenger, 114, 10. 
\bibitem{}
Boyarchuk A.A., Lyubimkov L.S., Sakhibullin N.A. 1985, Astrophysics, 22, 203.
\bibitem{}
Caffau E., Maiorca E., Bonifacio P., Faraggiana R., Steffen M., Ludwig H.-G., Kamp I., Busso M. 2009, A\&A, 498, 877. 
\bibitem{}
Caffau E., Ludwig H.-G., Bonifacio P., Faraggiana R., Steffen M., Freytag B., Kamp I., Ayres T.R. 2010, A\&A, 514, A92. 
\bibitem{}
Carlsson M. 1986, Uppsala Obs. Rep., 33, 1. 
\bibitem{}
Castelli F. 2005, Mem. Soc. Astron. Ital. Suppl., 8, 34. 
\bibitem{}
Castelli F., Kurucz R.L. 2003, in Piskunov N.E., Weiss W.W., Gray D.F., eds, Proc. IAU Symp. 210, Modeling of Stellar Atmospheres. Poster A20. Astron. Soc. Pac., San Francisco, P. A20. 
\bibitem{}
Claret A. 2004, A\&A, 424, 919. 
\bibitem{}
Cunto W., Mendoza C. 1992, Rev. Mexicana Astron. Astrofis., 23, 107. 
\bibitem{}
Cunto W., Mendoza C., Ochsenbein F., Zeippen C.J. 1993, A\&A, 275, L5.
\bibitem{}
Ekstr\"{o}m S., Georgy C., Eggenberger P., Meynet G., Mowlavi N., A. Wyttenbach A., Granada A., Decressin T., et al. 2012, A\&A, 537, A146. 
\bibitem{}
Fabbian D., Asplund M., Carlsson M., Kiselman D. 2006, A\&A, 458, 899. 
\bibitem{}
Heger A., Langer N. 2000, ApJ, 544, 1016. 
\bibitem{}
Heiter U., Barklem P., Fossati L. et al. 2008, Journal of Physics: Conf. Ser., 130, 012011. 
\bibitem{}
Hirata R., Horaguchi T. 1994, Atomic Spectral Line List. Department of Astronomy, Kyoto University, Kyoto. 
\bibitem{}
Korn A., Shi J., Gehren T. 2003, A\&A, 407, 691. 
\bibitem{}
Korotin S.A., Andrievsky S.M., Kostynchuk L.Yu. 1998, Astrophys. Space Sci., 260, 531. 
\bibitem{}
Kupka F., Piskunov N., Ryabchikova T.A., Stempels H.C., Weiss~W.W. 1999, A\&AS, 138, 119. 
\bibitem{}
Kurucz R. L., 1993, CD-ROM 13, ATLAS9: Stellar Atmosphere Programs and 2 km/s grid. Cambridge, Mass.: Smithsonian Astrophys. Obs. 
\bibitem{}
Kurucz R.L., Furenlid I., Brault J., Testerman L. 1984. Solar Flux Atlas from 296 to 1300 nm, New Mexico, National Solar Observatory. 
\bibitem{}
Luck R.E. 2014, AJ, 147, 137.
\bibitem{}
Luck R.E., Lambert D.L. 1985, ApJ, 298, 782.
\bibitem{}
Lyubimkov L.S. 1998, in Modern Problems of Stellar Evolution, ed. D.S. Wiebe, Moscow: Geos, p.231. 
\bibitem{}
Lyubimkov L.S. 2013, Astrophysics, 56, No. 4, 472. 
\bibitem{}
Lyubimkov L.S., Lambert D.L., Rostopchin S.I., Rachkovskaya T.M., Poklad D.B.    2010, MNRAS, 402, 1369 (Paper I). 
\bibitem{}
Lyubimkov L.S., Lambert D.L., Korotin S.A., Poklad D.B., Rachkovskaya T.M.,    Rostopchin S.I. 2011, MNRAS, 410, 1774 (Paper II). 
\bibitem{}
Lyubimkov L.S., Lambert D.L., Kaminsky B.M., Pavlenko Ya.V., Poklad D.B., Rachkovskaya T.M. 2012, MNRAS, 427, 11 (Paper III). 
\bibitem{}
Lyubimkov L.S., Lambert D.L., Poklad D.B., Rachkovskaya T.M., Rostopchin S.I. 2013, MNRAS, 428, 3497. 
\bibitem{}
Moore C.E. 1970, NSRDS-NBS 3, sec. 3. 
\bibitem{}
Rentzsch-Holm I. 1996, A\&A, 312, 966. 
\bibitem{}
Seaton, M.J. 1962, Proc. Phys. Soc., 79, 1005. 
\bibitem{}
Steenbock W., Holweger H. 1984, A\&A, 130, 319. 
\bibitem{}
St\"{u}renburg S., Holweger H. 1990, A\&A, 237, 125. 
\bibitem{}
Summers H.P. 2004, Atomic Data and Analysis Structure, User manual. University of Strathclyde (http://www.adas.ac.uk)
\bibitem{}
Suno H., Kato T. 2006, Atomic Data and Nuclear Data Tables 92, 407. 
\bibitem{}
Takeda Y. 1992, PASJ, 44, 649. 
\bibitem{}
Takeda Y., Honda S. 2005, PASJ, 57, 65. 
\bibitem{}
Tsymbal V.V. 1996, ASP Conf. Ser., 108, 198. 
\bibitem{}
Tull R.G., MacQueen P.J., Sneden C., Lambert D.L. 1995, PASP, 107, 251. 
\bibitem{}
van Leeuwen F. 2007, Hipparcos, the New Reduction of the Raw Data. Dordrecht; Springer. 
\bibitem{}
van Regemorter H. 1962, ApJ, 136, 906. 
\bibitem{}
Vernazza J. E., Avrett E. H., Loeser R., 1981, ApJS, 45, 635. 
\bibitem{}
Wilson N.J., Bell K.L., Hudson C.E. 2005, A\&A, 432, 731. 
\bibitem{}
Zatsarinny O., Bartschat K., Bandurina L., Gedeon V. 2005, PhRv A., 71, 42702. 


\end{thebibliography}
\end{document}